COMPUTATIONAL MODELING IN PYROMETALLURGY

# A First-Principles Tool to Discover New Pyrometallurgical Refining Options

MICHIEL J. VAN SETTEN ,[1,2,5] ANNELIES MALFLIET,[3] GEOFFROY HAUTIER ,[1,4] and BART BLANPAIN [3]

1.—Institute of Condensed Matter and Nanosciences, Université catholique de Louvain, Chemin des Étoiles 8, 1348 Louvain-la-Neuve, Belgium. 2.—IMEC, Kapeldreef 75, 3001 Leuven, Belgium. 3.—Department of Materials Engineering, Katholieke Universiteit Leuven, Kasteelpark Arenberg 44, 3001 Leuven, Belgium. 4.—Thayer School of Engineering, Dartmouth College, Hanover, NH 03755, USA. 5.—e-mail: michiel.vansetten@imec.be

We demonstrate the opportunities of first-principles density functional theory (DFT) calculations for the development of new metallurgical refining processes. As such, a methodology based on DFT calculations is developed to discover new pyrometallurgical refining processes that use the addition of a third element to remove an impurity from a molten host material. As a case study, this methodology is applied to the refining of lead. The proposed method predicts the existing refining routes as well as alternative processes. The most interesting candidate for the removal of arsenic from lead is experimentally verified, which confirms the suitability of the remover element. The method is therefore considered as a useful approach to speed up the discovery of new pyrometallurgical refining processes, as it provides an ordered set of interesting candidate remover elements.

## INTRODUCTION

The total amount of raw materials on our planet is limited. Although one could argue about the time frame at which they will be exhausted, in our current economy they will eventually. The only option to remedy this in the long run is to reach a truly circular economy.[1] Separating the raw materials in our waste streams, however, is not a simple task. Modern day electronics industry produces large volumes of products containing a significant portion of the elements from the periodic table, mixed in structures closing in on a scale of just a few nanometers. On top of this, the lifetime of these products is in general rather limited.[2,3] The recovery of purified metals from these heterogeneous secondary, but also from lower grade primary, sources requires more and more advanced metallurgical refining methods. The discovery of new pyrometallurgical refining processes, however, is difficult and complex.[4–6] A smart way of using computational approaches like density functional theory (DFT) may be used to speed up this discovery process.

Since the foundation of DFT in 1964,[7,8] a growing range of applications in physics, chemistry, and materials science have emerged which apply first-principles calculations to predict and understand the behavior of complex systems. In materials science, the ability of first-principles calculations to predict material properties has been applied among others to understand interface and diffusion phenomena, to model phase equilibriums, and to predict mechanical properties.[9] The progress in computational speed and in robustness of the algorithms has made high-throughput first-principles calculations possible, which allow one to screen a broad range of material compositions on their properties and to establish large databases for data-mining applications.[10] High-throughput calculations can be especially useful to limit the amount of experimental efforts. Although thermodynamic databases based on first-principles calculations are available,[11,12] their application to develop processes is limited. An exception is, for example, the work of Jain et al.[13] in which 22 pure metals were screened with DFT calculations to find a suitable element to remove Hg from syn-gas streams at high temperature. By applying relatively simple bulk calculations using structural data available in online databases, it was possible to qualitatively predict the trends in the few available experimental data. This study illustrates the ability of first-principles-based screening methods to guide the focus of future experimental plans, thereby accelerating the pace at which processes could be developed.

In this paper, this methodology is further explored to evaluate the strength of first-principles calculations to search for new pyrometallurgical refining methods. In a metallurgical flow-sheet, the refining process is intended to remove the impurity elements up to the desired level. Removal of the impurity elements from the base metal can improve the properties of the product and can create added value in the case of recovery of valuable by-products. Both pyrometallurgical and hydro-metallurgical refining routes exist, and many of them already have a long history. It was, however, only in the middle of the twentieth century that chemical thermodynamics were applied to the metallurgical field, which led to the transition of metallurgy from a workmanship or skill to a science. The introduction of thermodynamics into the metallurgical field allowed the studying of the process steps as chemical reactions between species, thereby providing among others data on solubility limits, the selectivity of different reactions, and their changes in enthalpy. For example, the refining method in which a third element (Rem) is added to a molten bath (Host) to remove an impurity (Imp) through the formation of an Imp–Rem compound can be described by the chemical reaction:

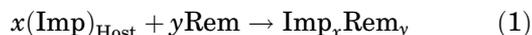

$$x(\text{Imp})_{\text{Host}} + y\text{Rem} \rightarrow \text{Imp}_x\text{Rem}_y \quad (1)$$

for which the thermodynamics determine the direction and extent of the reaction. It is for this reaction that, in this paper, a method is developed in which first-principles calculated thermodynamic data are used as a screening method for the development of new metallurgical processes. The advantage of the proposed first-principles method is that it makes use of the available existing data and computational power to search for possible remover elements Rem from a large set of elements. It allows the easy finding of other, maybe more effective, candidate remover elements Rem, for which only a limited set of experiments are then needed to verify the predicted refining ability.

The first-principles method developed in this paper is applied to the refining of lead as a case study. Typical impurity elements in primary or secondary Pb resources are Ag, As, Bi, Cu, Sb, and Sn. The refining of Pb is performed in kettles and furnaces to remove the impurities in several steps.[14] Firstly, the limited solubility of Cu in molten Pb is used to remove this element. Molten Pb is cooled below the freezing point of copper to 340–350°C, after which the crystallized copper, also containing, besides Cu, any remaining Ni, Co, and Zn, can be removed by skimming. After removal of this dross, the Cu content can be further decreased by the addition of a sulfur source, leading to the formation of copper sulfides.[15] The molten Pb is then transferred to a reverberatory softening furnace where the temperature is raised again to 700–750°C and air is blown into the vessel. The Sn, As, and Sb are consequently oxidized, and these oxides are skimmed off. As an alternative to the reverberatory furnace process, a mixture of sodium hydroxide and sodium nitrate is stirred into a kettle containing the molten Pb. The melt is maintained at 450–500°C to remove Sn, As, and Sb, as sodium stannate, arsenate, and antimonate, in the flux on top of the lead bath. Zn is removed as ZnO. This is called the Harris process, which can also be used to recover In.[16] After softening, the next step is the removal of Ag. This is performed in desilvering kettles in which Zn is added as the remover element. Zn will form, with Ag and also with Au, inter-metallic compounds, which form as a crust on top of the lead bath upon cooling below 370°C, but above the melting temperature of Pb. This crust can then be removed and treated for further recovery of the precious metals. The excess Zn is removed from the molten Pb by reheating the Pb under vacuum conditions to 500°C to make the Zn more volatile. Finally, the Betterton–Kroll[17,18] process can be used to decrease the Bi content. In this process, Ca and Mg are added as remover elements to the softened, desilvered, and dezinced Pb. Bi reacts with Ca and Mg to form $CaMg_2Bi_2$, which can be skimmed off the melt. Achievable levels of Bi are below 50 ppm, or even 10 ppm, when Na is also added. If detellurizing is required, it can be done prior to either softening or desilverizing. Detellurizing prevents Te contamination of the silver dross, from which it can only be removed with great difficulty and expense. A possible flowsheet for detellurizing is described in the work of Garbuzov in 1960, as described in the review on Te removal from lead- and copper-bearing materials by Makuei and Senanayake.[19] After removal of copper from the molten Pb, metallic Na is added as a lead-sodium alloy. Na reacts with Te to form $Na_2Te$, which floats as a solid compound to the surface of the molten Pb. A molten sodium hydroxide layer is used to collect the sodium telluride, and this dross is then removed and is suitable for the recovery of Te.

With respect to the aim of this study, to verify the ability of the proposed first-principles method to find suitable remover elements Rem the lead case study is interesting as the results can be compared with the Parkes and Betterton–Kroll process, in which this refining method is applied to remove, respectively, Ag by Zn and Bi by Ca/Mg. Also, the ability of Na to remove Te should follow from the first-principles calculations. Besides confirmation of the existing processes, the first-principles method is used to compare the efficiency of the current refiner elements with alternative elements.

## METHODOLOGY

### Development of the Novel Approach

The first-principles method in this paper is developed for Eq. 1. Assuming a constant temperature during this reaction, the change in Gibbs free energy of this reaction can be written as:

$$\Delta G = \Delta H - T\Delta S \quad (2)$$

The change in enthalpy $\Delta H$ results from the change in enthalpy between the products and reactants, and from the change in mixing enthalpy. The entropy term $\Delta S$ can have three sources: configurational, electronic, or vibrational. For a negative change in free energy of the system, $\Delta G < 0$, the refining Eq. (1) will proceed towards the formation of Imp–Rem compounds, thereby removing Imp from the Host melt. When $\Delta G = 0$, the equilibrium concentration of Imp in the melt is reached. The strategy to find new pyrometallurgical refining methods is based on Eq. 2 and consists of five steps.

*Find Elements Rem Which Could Refine the Melt Through Eq. 1*

The first criterion for possible remover elements Rem for the refining Eq. 1 is that they react with the impurity Imp in the melt to form a stable compound. This is approached in this study by considering, for Eq. 2, that the formation enthalpy of the compound is sufficiently large to overcome the loss in configurational entropy caused by the purification of the melt at the melting temperature of the Host:

$$\Delta H_{f,ImpRem_x} > T_m \Delta S \quad (3)$$

This criterion neglects the enthalpy change of mixing and the electronic and vibrational contribution to the entropy in Eq. 2. A database of precomputed first-principles formation enthalpies is queried to find values for the formation enthalpies of the Imp–Rem compounds in Eq. 3. For the configurational entropy in this equation, the expression for an ideal solution of Imp in the Host is applied:

$$-\frac{S_{conf}}{k_B} = N_1 \ln(N_1/N) + N_2 \ln(N_2/N) \\ + (N - N_1 - N_2) \ln((N - N_1 - N_2)/N) \quad (4)$$

For the purification Eq. (1), $N_1$ and $N_2$ are the amounts of impurity and remover atoms, respectively, which follow directly from Boltzmann's equation for the entropy change and Stirling's approximation for the factorial of a large integers. The ratio between them, $1/x$, is determined by the stoichiometry of the impurity–remover compound that is expected to form. For the lead refining case in this study, it is considered that the formation enthalpy must overcome the ideal entropy of mixing of 1 in 1000 atoms at the melting temperature of Pb. This selection of 1 in 1000 atoms is arbitrarily chosen here as a first screening to eliminate those elements Rem that form stable compounds with Imp but would lead to low refining levels.

The approach taken here is solely based on the thermodynamics of Eq. 1 and the entropy originating from the mixing. Any kinetic barriers are for now neglected. However, since the sought-after reactions take place in a bath of molten metal, it seems to be a fair assumption that, in general, sufficient kinetic energy is available to overcome most kinetic barriers. In the case of high melting points of remover elements and/or low solubility limits, the refining process, however, may take relatively long to complete.

*Determine the Stability of the Imp–Rem Compound in the Host–Imp–Rem System*

A second criterion imposed on the possible remover elements remaining from step 1 is that the remover elements Rem should preferentially react with the impurity Imp compared to reacting with the Host. This criterion is based on the fact that similarly to the refining reaction (1), a competing reaction could occur in which Rem reacts with the Host to form a Host–Rem compound:

$$(\text{Host})_{Host} + y\text{Rem} \rightarrow \text{HostRem}_y \quad (5)$$

This criterion that for elements Rem the Reaction (1) should be energetically favorable compared to Reaction (5), is approached in this study by comparing the enthalpies of formation of the Imp–Rem to the Host–Rem compound. Using solely precomputed enthalpies of formation, the ternary Host–Imp–Rem phase diagram at 0 K is constructed. Only those elements are retained for which the actual formation of the Imp–Rem compound is indicated by the ternary phase diagram. Those elements for which the formation of Host–Rem compounds is predicted are removed from the list.

*Predict the Impurity Concentration in the Host Metal*

Next, the impurity concentration that can be reached at a given temperature is predicted using the formation enthalpy of the Imp–Rem compound and the configurational entropy of Imp in the Host. For this, Eq. 3 is used with an equality sign to find the value of $N_1$ for which the entropy term would equal the enthalpy term. For lower values of $N_1$, the formation enthalpy of the Imp–Rem compound is insufficient to overcome the loss in entropy of the melt. With this information, a ranking of the remover elements for a given Host and Imp can be made, also taking into account economic and environmental constraints.

In our approach, we compare the formation enthalpies of Imp–Rem compounds at 0 K to the ideal entropy of mixing to determine if a purification reaction is likely. By doing this, we make the approximation that the enthalpy and entropy change from 0 K to the molten liquid phase cancels between the compounds of the phase diagram. For various combinations of hosts and impurities, however, oxygen appears as a predicted remover element. Considering that oxygen should occur in the gas phase would break this cancellation. However, since, in the known methods where oxygen is used in refining, oxygen is blown into the mixture, the reference state of oxygen in the gas phase is not the most relevant. Treating oxygen just as the other elements, we also recover the known oxygen refining processes.

*Improve the First-Principles Calculations*

For selected cases of higher importance, the first-principles results that enter the construction of the phase diagram can be recalculated at higher precision. In addition, data for known compounds not present in the database, and results from super-cell calculations pertaining to low concentrations of the impurity and remover element in the host metal, can be added.

*Validation of the First-Principles Results Through Experiments*

Finally, the new refining method can be experimentally validated for the most promising remover elements for each impurity.

**Implementation of the Methodology**

*Selection of the System*

Within this study, the new methodology was applied to the system with Pb as the host metal, containing the impurities As, Te, Sb, Sn, In, Cu, and Ag. Although in practice these impurities can simultaneously be present in the melt, and a remover element R could lead to reactions involving more than one impurity, it is assumed in this study that only one impurity is present in the melt.

*Data Mining for Remover Elements*

The database of the materials project[11] was used to obtain the initial set of possible remover elements and for the construction of the corresponding phase diagrams. This database contains first-principles results of DFT calculations performed using the Perdew–Burke–Ernzerhof (PBE) generalized gradient approximation (GGA) exchange correlation functional[20,21] on all materials known in the Inorganic Crystal Structure Database (ICSD) containing less than 100 atoms in the unit cell. The PBE functional is still considered the most transferable GGA, and will be used in all calculations in this work. All calculations were performed spin-polarized, and a parameter U, representing the strengths of the on-site coulomb interaction, was used for selected elements. The database queries were performed using python tools based on the pymatgen package[11] inside jupyter notebooks. For step 1, a stability threshold of 0.8 eV per atom was used. This value is needed to overcome the ideal entropy of mixing of 1 in 1000 atoms at the melting temperature of Pb.

*Computational Details*

To increase the precision of the computations on the most promising systems, i.e., systems of remover elements that are environmentally and economically viable and promise a high purification level, tighter convergence parameters were employed. The k-point density is especially important, as metals are often involved in the reactions under investigation. In these higher precision recalculations, it was aimed to stay as close as possible to the original computational settings in the materials project. They were hence performed with the VASP package using the PBE exchange correlation functional.[20–24] The k-point density was increased to 10,000 k-points per reciprocal atom, and a Gaussian smearing with a 0.05 eV broadening was applied. In addition, the stopping criteria on the electronic and structural relaxation were both decreased by one order of magnitude, $1 \times 10^{-5}$ eV and $1 \times 10^{-4}$ eV, respectively.

*Experimental Validation*

The ability of the new methodology to correctly predict suitable remover elements to refine a specific impurity from the host metal was experimentally validated. For the given case study of Pb, As was chosen as the impurity element. 1.5–5 g of Pb (99.99% purity) was mixed with 0.5–2 wt% of As and a certain quantity of the remover element Rem. This quantity was taken hyperstoichiometric compared to the stoichiometry of the compound which was predicted by the ab initio calculations to form between As and Rem. This mixture was sealed in a vacuum quartz tube after flushing three times with Ar to avoid O contamination. The quartz tube and its content were then heated in a vertical tube furnace to 1000°C for 12 h to allow the impurity and remover element to dissolve and to react. The quartz tube was then dropped inside water to cool down the sample. One mixture was removed from the quartz tube for investigation, whereas one mixture was given an extra heat treatment at 400°C for 24 h inside the quartz tube to verify the predicted effect of temperature on the refining ability. After removal from the quartz tubes, the Pb samples were embedded in resin, ground and polished for electron probe micro-analysis (EPMA) using wavelength dispersive spectroscopy (WDS). The embedding was carried out using Technovit 4004 resin. For the grinding, 800-, 1200-, and 4000-grit SiC grinding papers were used. To avoid the pressing of the SiC particles into the soft Pb matrix, paraffin wax was first applied on the grinding papers. Polishing cloths with diamond paste of, respectively, 3 m and 1 m was used. The EPMA-WDS analysis was performed on a FEG-EPMA JXA-8530F from JEOL, operated at 15 kV and 30 nA. The composition of the formed compounds and the remaining As content in the Pb matrix were measured to compare them with the results of the calculations.

## RESULTS AND DISCUSSION

The methodology described in Sect. "Development of the novel approach" section was applied to the system with Pb as the host metal, containing the impurities As, Te, Sb, Sn, In, Cu, and Ag. For all the impurities, steps 1 to 3 were performed to identify the most interesting remover elements. Only for As and Te, was step 4 of Sect. "Development of the novel approach" section also executed. The experimental validation (step 5 of Sect. "Development of the novel approach" section), was performed for As and its most promising remover element based on the achievable level of refinement.

### Predicted Candidate Remover Elements

Table I lists the results of the database search and additional first-principles calculations for the most promising remover elements for As, Te, Sb, Sn, In, Cu, and Ag as impurities in Pb. Y appears as the most efficient remover element for As, Sb, and Bi. Ba, Ca, and Sr are suitable to remove Te to low levels. Sn and In form stable oxides. For Cu and Ag as impurities in Pb, no suitable remover elements were found.

### Evaluation of the Predicted Remover Elements

For In, Sn, As, Sb, Bi, and Te, the possible elements Rem to refine Pb as predicted by the first-principles calculations are compared in Fig. 1.

For Sn and In, O is the most suitable remover element. For Sn, this is also what is industrially applied by blowing air into the reverberatory furnace in which first Sn will oxidize. Based on the first-principles calculations, this will Pb to the formation of SnO and consequently to $Sn_5O_6$.

During the blowing of air in to the reverberatory furnace, As and Sb are also removed. For the removal of these elements in the softening process, the first-principles calculations predict that the addition of O to the molten Pb causes the formation of lead arsenates and lead antimonates, which agrees with the losses of Pb to the dross. The first-principles calculations, however, also indicate a more efficient remover element for As and Sb. For both elements, the addition of Y as remover element would lead to lower refining levels compared to O. In addition, it would lead to a pure Imp–Rem compound instead of ternary compounds containing Pb. This would decrease Pb losses to the dross.

For the removal of Bi by the Kroll–Betterton[17,18] process by the addition of Ca and Mg, the suitability of Ca as remover element is confirmed by the first-principles calculations. Ba, Sr, Ca, and Mg are all alkaline earth elements (second column in the periodic table) for which their efficiency to remove Bi decreases with decreasing atomic number. Mg is

Table I. Most promising remover elements and refining level predicted from the first-principles calculated formation energies of the Imp–Rem compounds using 0.8 eV as a stability threshold

| Impurity | Removers (estimated fraction of impurity (PPM) at 600 K) |
|---|---|
| As | Y(0) Sc(2) Zr(16) Sr(71) O(90) Ca(223) |
| Te | Ba(0) Ca(0) Sr(0) Sc(1) Y(5) |
| Sb | Y(22) Th(64) O(69) Ba(167) |
| Bi | Y(430) |
| Sn | O(2) |
| In | O(0) |
| Ag | |
| Cu | |

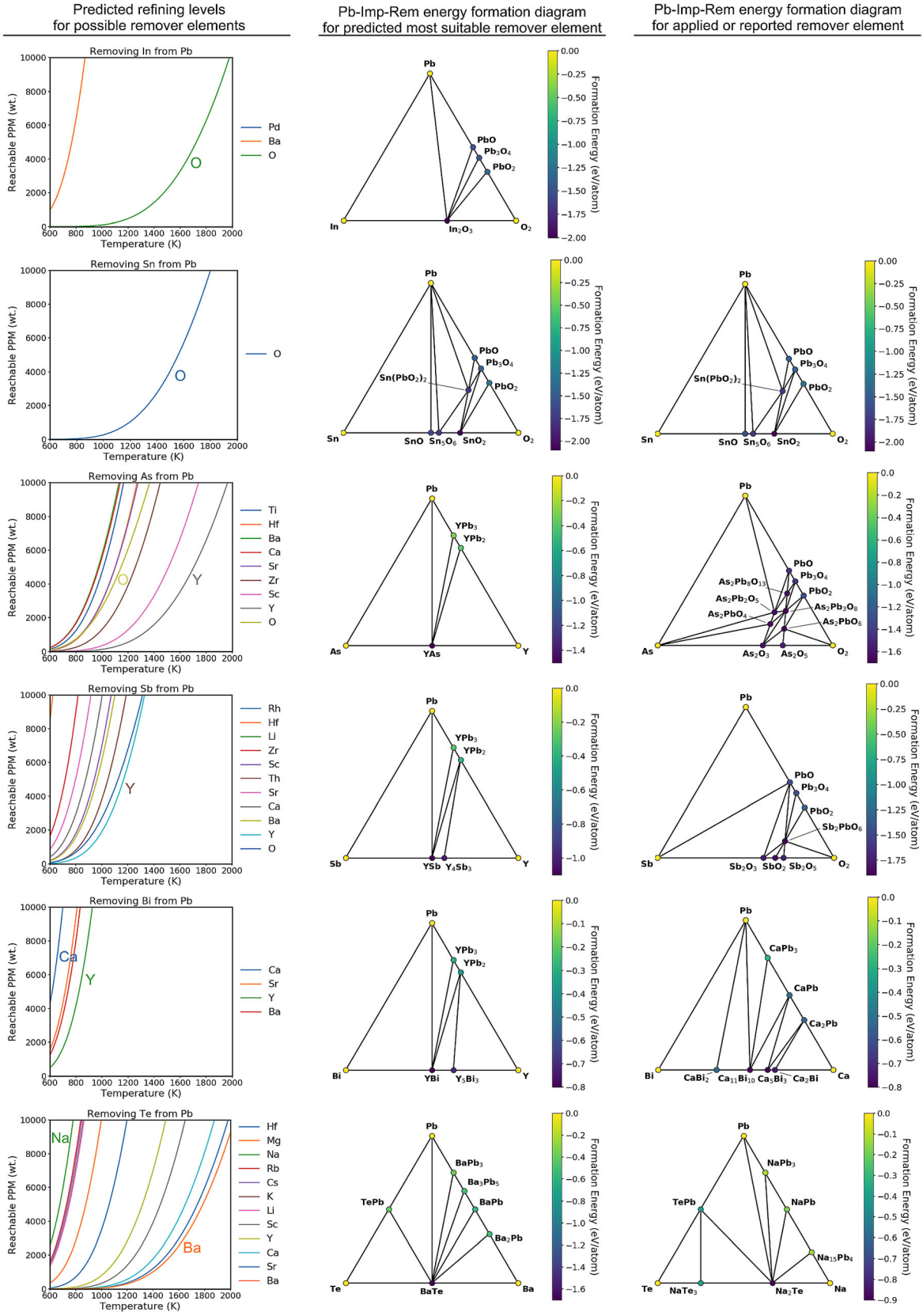

Fig. 1. Predicted refining levels for the possible remover elements for In, Sn, As, Sb, Bi, and Te. For the most suitable remover element, the energy of formation diagram is compared to the one for an applied or reported remover element. The formation diagram axis represent mole fractions and the formation energies are relative to the pure elements.

A First Principles Tool to Discover New Pyrometallurgical Refining Options

not considered in the graph in Fig. 1, as the formation energy of $Mg_2Bi_3$ is below the considered stability threshold. The combined effect of adding both Ca and Mg, as is industrially applied, is, however, not considered by the current calculations. Including a combination of multiple elements is in principle also computationally possible but beyond the scope of this work. In the case of adding only a single remover element, Y appears again to be the most efficient remover element. The formation energy of the binary Im-Y compounds, however, decreases from As to Sb and Bi, which is also their order in the fifth column of the periodic table with increasing atomic number.

The removal of Te is reported to be possible by the addition of Na through the formation of the intermetallic compound, $Na_2Te$, which is effectively predicted by the first-principles calculations as the compound which would form upon adding Na to a Pb melt containing Te. The possible refining level of Na is, however, low compared to several other elements, according to the predictions. Ba is the most efficient remover element, but Y also appears again as a suitable remover element due to the formation of YTe.

Despite the agreements between the actual and predicted abilities for certain impurity-remover elements, the first-principles calculations do not indicate the ability of S to remove Cu and of Zn to remove Ag. The calculated ternary Pb-Cu-S and Pb-Ag-Zn phase diagrams are shown in Figs. 2 and 3, respectively. We speculate that, in this case, the approximation of canceling solid to liquid entropy changes may be violated.

Zn is industrially applied to remove Ag through the formation of Zn-Ag compounds, whereas Zn is not predicted as a possible remover element for Ag. The reason for the absence of Zn, or any other candidate remover element, in Table I is that the formation energy of the Ag-Rem compounds are all lower than the considered threshold of 0.8 eV. This threshold level is the energy needed to overcome the configurational entropy to remove 1 impurity atom out of 1000 atoms at the melting temperature of Pb. The first-principles calculations, however, effectively predict the stability of the $ZnAg_3$ compound compared to the stability of a Pb-Zn compound (Fig. 2). However, the low formation energy of the $ZnAg_3$ compound ($< 0.1$ eV) is not in agreement with the actual refining levels of Ag using Zn as remover element, which is reported to be below 10 ppm.[16] A possible explanation is that the Parkes process for desilverization is in fact a liquid–liquid extraction process. Molten Zn and molten Pb form two immiscible liquids, for which Ag has a 3000 times higher solubility in molten Zn than in molten Pb. Ag therefore prefers to migrate to the molten Zn which is initially stirred into the molten Pb but will eventually separate to the surface. It is upon cooling that the intermetallic Ag-Zn compounds then form a crust on top of the molten Pb pool. Therefore, it is the difference in solubility of Ag in molten Zn and molten Pb that determines the attainable refining levels, rather than the stability of the formed intermetallic compound which the first-principles method uses to predict the refining level. The ability of Zn to remove Ag from Pb is therefore not predicted by the applied first-principles method.

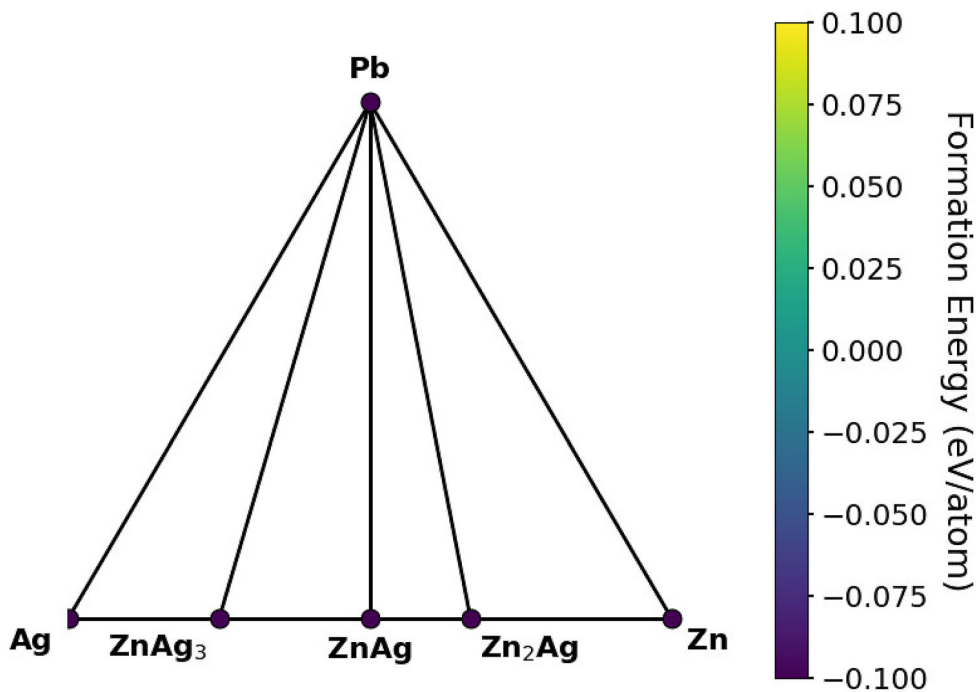

Fig. 2. Pb-Ag-Zn phase diagram with the first-principles calculated formation energies. Ag-Zn compounds preferentially form in the ternary Pb-Ag-Zn system, but with a low energy of formation. The formation diagram axis represent mole fractions and the formation energies are relative to the pure elements.

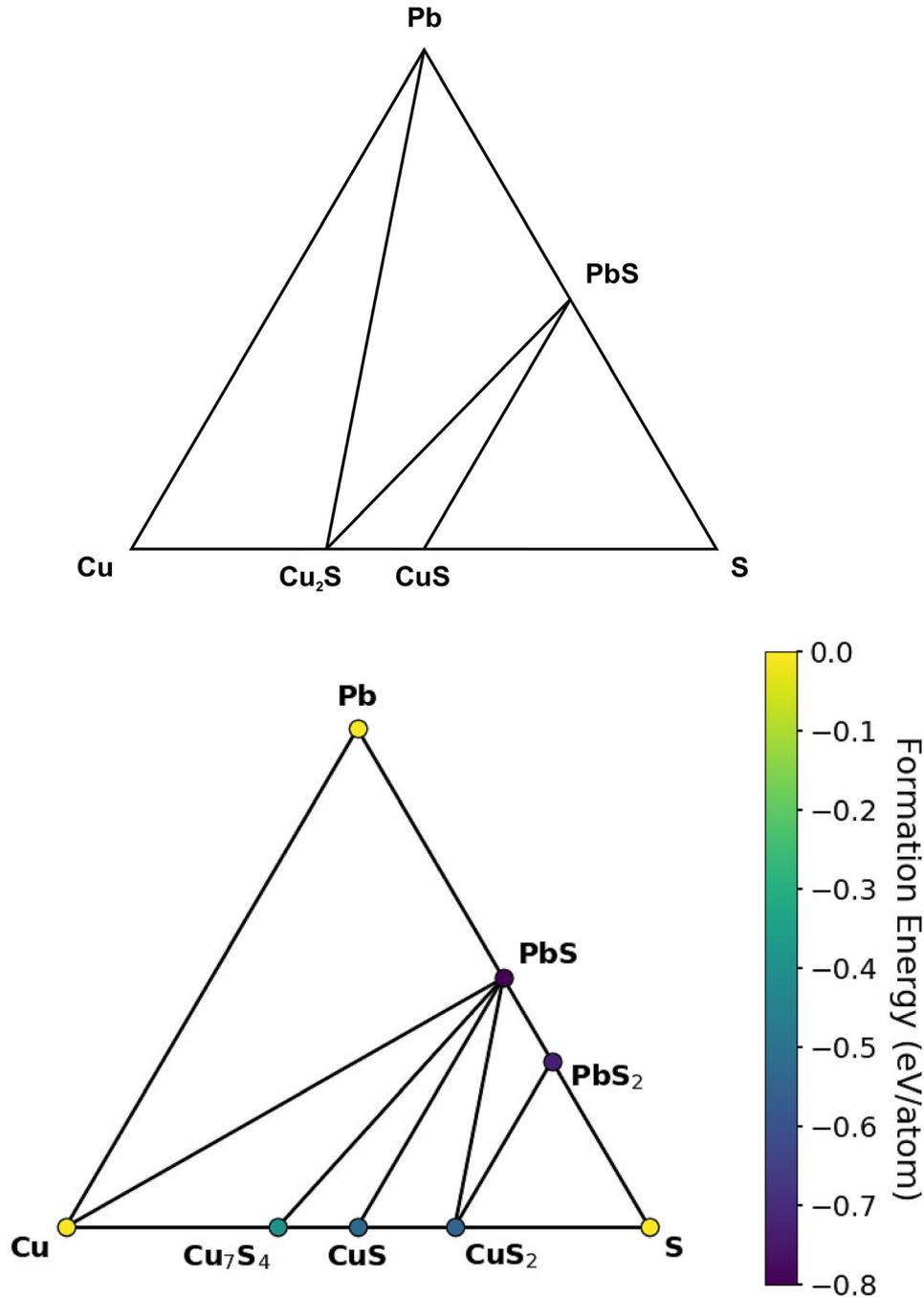

Fig. 3. Pb-Cu-S phase diagrams according to Meissner[25] and as calculated with first-principles, including the calculated formation energies. The preferential formation of PbS compared to a Cu-S compound is not predicted by the first-principles calculations. The formation diagram axis represent mole fractions and the formation energies are relative to the pure elements.

For Cu, the first-principles calculated phase diagram is compared with the accepted phase relationships in the Pb-Cu-S system in Fig. 3. In the calculated phase diagram, the stability of $Cu_2S$ in the binary Cu-S system is not predicted, and as such its preferred stability compared to PbS is also not predicted. The reason for this is unclear. Nonetheless, it is interesting to note here that there is also a discrepancy between the practical ability of sulfur to remove Cu and the level of refining that can be expected from the thermodynamic properties in the Pb-Cu-S system. In this respect, Friedrich et al.[15] discussed the influence of alloying elements, such as Sn and As, on the reaction mechanisms and kinetics

A First Principles Tool to Discover New Pyrometallurgical Refining Options

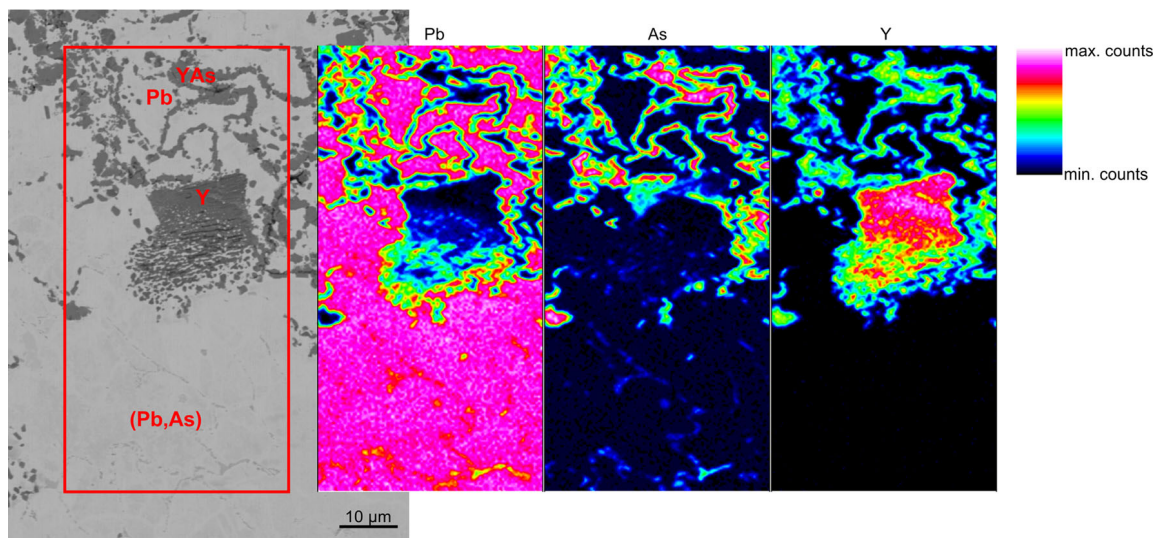

Fig. 4. Backscattered electron image and WDS elemental mappings of Pb, As, and Y. The formation of YAs in the top part of the image confirms the first-principles predictions.

occurring during the removal of Cu, which can explain the lower achievable levels of refining than thermodynamically predicted.

**Experimental Verification**

Due to the predicted high efficiency of Y to remove As, Sb and Bi from Pb, the refining ability of Y in the Pb-As system was experimentally verified. The microstructure of the sample with a hyperstoichiometric ratio Y:As of 1.1:1 after a heat treatment of 12 h at 1000°C is shown in Fig. 4. The sample consists of two type of zones. In one zone (in the upper part of the BSE image), Y has reacted with As to form YAs, as predicted by the first-principles calculations. The Pb matrix in between these particles is depleted in As from the original concentration of 2 wt% to $0.04 \pm 0.01$ wt% As ($400 \pm 100$ ppm). Due to the high melting point of Y and its low solubility in Pb, not all Y has so far reacted with As. This can be seen in the center of the BSE image in Fig. 4, where a dissolving and reacting yttrium particle is still visible. This is the reason why, in the other zone (in the lower part of the BSE image), the Pb matrix still contains As. This As has segregated to the Pb grain boundaries during cooling. To verify the refining ability of Y, a sample with a higher ratio Y:As of 5:1 was first heat-treated at 1000°C for 12 h, followed by a heat treatment of 24 h at 400°C. Both the excess Y and the decreased temperature should enhance the removal of As by Y. The result is shown in Fig. 5. The excess Y has caused the formation of an Pb-Y compound besides the YAs particles. In the Pb matrix between the particles, no segregation of As was observed. In the WDS spectra in Fig. 5, the presence of Y in the Pb matrix has been identified, but for As no peak was observed. In the spectrometer range in which the presence of As should give a peak in intensity, only a smooth background signal was detected. Using quantitative spot analyses, the measured concentration of As in the Pb matrix has been measured as $0.002 \pm 0.001$ wt% As ($20 \pm 10$ ppm). The predicted and experimentally determined refining ability of Y for As are in reasonable agreement (Fig. 6).

**Evaluation of First-Principles Calculations to Find New Refining Routes**

For the refining methods based on the addition of a third element to a metal containing one impurity element, the performed first-principles calculations are in agreement with existing practices or data reported in literature. In addition, the proposed methodology proved to be an efficient way to find other remover elements. With a limited series of experiments in the Host–Imp system, the suitability of the candidate remover elements, R, was easily verified. In addition, the systematic approach of calculating the formation energy of Imp–Rem compounds for a large part of the periodic table allows one to find trends in to which degree impurity elements from a certain column like to react with a certain remover element, as illustrated in Fig. 1, by the ability of Y to refine, respectively, As, Sb, and Bi. It can also be helpful to define the correct procedure for which elements to refine first.

Of course, the currently performed first-principles calculations in the Host–Imp–Rem systems do not fully represent the complex industrial situation. The melt typically contains more than one impurity element as assumed in the first-principles calculations. Also, other phenomena are not included in the currently performed calculations, such as the remover element forming a liquid immiscibility gap with the melt, as is the case of the addition of Zn to the liquid Pb melt. It is therefore suggested to combine the first-principles calculations with experiments (on the laboratory or pilot scale), in which the calculations can be used as a novel tool to design the most interesting refining experiments. Besides verification of the calculations, the experimental results can in addition already give insights into the kinetics of the process. For the experimental tests in this study, for example, Y has a high melting point and a low solubility in liquid Pb. This leads to the slow dissolution kinetics of the yttrium particles (Fig. 4), causing incomplete formation of YAs even after 12 h at 1000°C (without stirring).

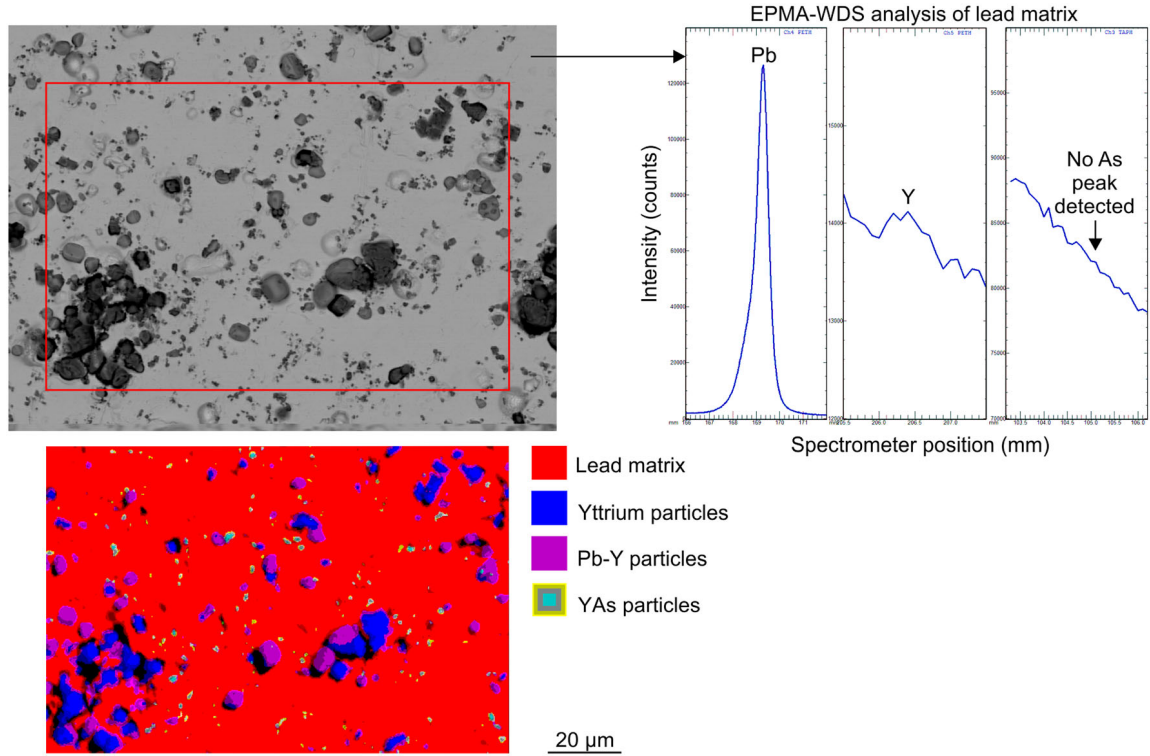

Fig. 5. BSE image and phase distribution map (as constructed from WDS elemental maps) of the Pb-As-Y sample in which the excess Y has caused the formation of Pb-Y compounds besides the YAs particles. The WDS spectra measured on the Pb matrix show that the level of As has decreased to this extent that it could not be detected anymore.

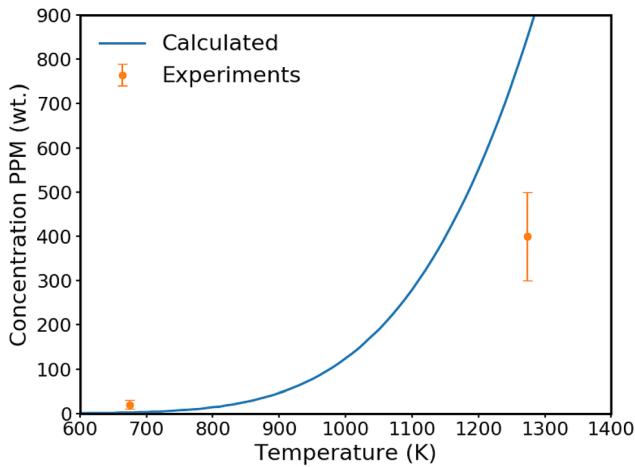

Fig. 6. Removing As from Pb by Y, comparing the predicted reachable concentration (black line) to the experimentally achieved values (red dots) (Color figure online)

To conclude, the value of the first-principles calculations is to provide the set and order of candidate remover elements, for which the appropriate laboratory- and industrial-scale tests can then be designed to evaluate which refining method is the most effective and efficient, and also economically the most interesting, to refine the impurities up to their desired level.

A possible disadvantage of this method is that new refining routes can only be found if the corresponding phase diagrams are sufficiently complete. If the stable phase between the impurity and the remover element is not present in the database, that specific route cannot be predicted. The method will hence work best for material systems in which the first-principles data are relatively complete. It is therefore important to keep these databases growing.

## CONCLUSION

A novel approach to search for new pyrometallurgical refining methods has been developed. In this method, precomputed first-principles formation enthalpies were used to identify the set of third elements that bind more strongly with the impurity element compared to the host metal. For those elements that bind strongly enough with the impurity to overcome the configurational entropy of the impurity in the host metal, ternary phase diagrams were constructed to confirm the actual formation of the expected compound. Only for the most interesting remover elements are experiments then needed to confirm their ability to remove the impurity element. This approach has been applied to the refining of Pb. The calculated results in general agree with respect to the current practice and knowledge of refining processes using a third element as a remover element, but they in addition suggest alternative remover elements that could provide better refining levels. For one of the suggested remover elements, the refining ability was experimentally verified. The predicted impurity-remover compound effectively preferentially formed compared to the remover–host compound. The proposed methodology based on first-principles calculations therefore provides a useful tool in the design of pyrometallurgical refining routes.

## ACKNOWLEDGEMENTS

Computational resources have been provided by the supercomputing facilities of the Université catholique de Louvain (CISM/UCL) and the Consortium des Equipements de Calcul Intensif en Fédération Wallonie Bruxelles (CECI) funded by the Fonds de la Recherche Scientifique de Belgique (FRS-FNRS). Funding via the Agency for Innovation and Entrepreneurship (VLAIO) project HBC.2016.0733 of the Flemish region with Campine is acknowledged.

## CONFLICT OF INTEREST

On behalf of all authors, the corresponding author states that there is no conflict of interest.